\documentclass[aps,prd,eqsecnum,preprint,tightenlines,nofootinbib,showpacs]{revtex4-1}
\usepackage{graphicx,amsmath,latexsym}

\def\bea{\begin{eqnarray}}
\def\eea{\end{eqnarray}}
\def\be{\begin{equation}}
\def\ee{\end{equation}}
\newcommand{\ub}[1]{\underline{#1}}
\newcommand{\ob}[1]{\overline{#1}}
\newcommand{\Pminus}{{\cal P}^-}

\begin{document}

\title{A light-front coupled-cluster method \\
for the nonperturbative solution \\
of quantum field theories}

\author{Sophia S. Chabysheva}
\author{John R. Hiller}
\affiliation{Department of Physics \\
University of Minnesota-Duluth \\
Duluth, Minnesota 55812}

\date{\today}

\begin{abstract}
We propose a new method for the nonperturbative
solution of quantum field theories and illustrate
its use in the context of a light-front analog
to the Greenberg--Schweber model.
The method is based on light-front quantization
and uses the exponential-operator technique of the
many-body coupled-cluster method.
The formulation produces an effective Hamiltonian
eigenvalue problem in the valence Fock sector of
the system of interest, combined with nonlinear
integral equations to be solved for the functions
that define the effective Hamiltonian.  The method
avoids the Fock-space truncations usually used in
nonperturbative light-front Hamiltonian methods and, 
therefore, does not suffer from the spectator dependence,
Fock-sector dependence,
and uncanceled divergences caused by such truncations.
\end{abstract}

\maketitle

\section{Introduction}
\label{sec:Introduction}

The central problem of a quantum field theory is to
compute its mass spectrum and the corresponding
eigenstates.  All physical quantities can be computed
from these.  If the theory is quantized in terms of
light-front coordinates~\cite{Dirac}, this spectral
problem can be written as a Hamiltonian eigenvalue
problem~\cite{DLCQreview}, 
${\cal P}^\mu|\psi(\ub{P})\rangle=P^\mu|\psi(\ub{P})\rangle$,
where $\Pminus\equiv{\cal P}^0-{\cal P}^z$ is the light-front
energy operator, 
$\ub{\cal P}\equiv({\cal P}^+={\cal P}^0+{\cal P}^z,
\vec{\cal P}_\perp=({\cal P}^x,{\cal P}^y))$ is the light-front
momentum operator, and $P^\mu$ are the corresponding eigenvalues.
For an eigenstate of mass $M$, the mass-shell condition
$P^2=M^2$ yields $P^-=(M^2+P_\perp^2)/P^+$.
Thus, eigenvalues of $\Pminus$ determine the mass spectrum.

The standard light-front Hamiltonian approach is to
expand $|\psi(\ub{P})\rangle$ in a set of Fock states, 
eigenstates of $\ub{\cal P}$ with definite numbers
of constituents.  The coefficients in the expansion
are the light-front momentum-space wave functions.  This
takes advantage of two important aspects of light-front
coordinates~\cite{DLCQreview}: the relative-momentum
coordinates separate from the external momentum, so that
the wave functions depend only on the relative momenta,
and the positivity of $P^+=\sqrt{(\vec P)^2+M^2}+P^z$
excludes vacuum contributions to the expansion, so that
the wave functions represent the properties of the
eigenstate only.

Given the Fock-state expansion, the eigenvalue problem
becomes an infinite set of coupled integral equations for
the wave functions.  The expansion and the
coupled system are truncated to yield a finite problem,
which is then solved, usually by numerical
techniques~\cite{TwoPhotonQED}.

In more than two dimensions, some form of regularization
is required to properly define the integrals of the
coupled system.  The cancellations that must take place in the
regularization scheme are disrupted by the truncation, 
resulting in uncanceled divergences.  A re-parameterization
of the theory, such as sector-dependent 
parameterization~\cite{SectorDependent,hb,Karmanov,Vary,SecDep}, 
can be arranged to
appear to absorb these divergences, but not simultaneously
for all physical quantities~\cite{SecDep}.  The truncation
also causes self-energy contributions and vertex functions
to be dependent on the momenta of Fock-state constituents
that are only spectators to the process in question.  This
spectator and Fock-state dependence results in great
complications for the analysis and solution of the theory.

\begin{figure}[hb]
\vspace{0.2in}
\centerline{\includegraphics[width=10cm]{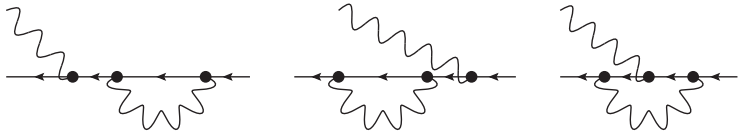}}
\caption{\label{fig:Fig1} Graphs contributing to
the Ward identity in QED.  Only the first contributes
in a one-photon truncation of the Fock space.}
\end{figure}

In particular, the Ward identity of gauge theories is
destroyed by truncation.  For photon emission in QED,
a one-photon truncation keeps only the self-energy
correction to the electron leg on the side opposite
the photon emission; the self-energy correction
on the other leg and the vertex loop correction 
are eliminated~\cite{OSUQED,hb}.  The relevant
diagrams are shown in Fig.~\ref{fig:Fig1}; only the first
survives the truncation.  
Thus, the Ward identity connecting
vertex and wave-function renormalization is broken for
interactions internal to a bound-state problem.
This is what drives the renormalization of the charge in a
sector-dependent parameterization of the theory~\cite{hb,Karmanov},
but this is clearly unphysical and has nothing to do with
ordinary charge renormalization.\footnote{For external photon
emission, the truncation
does not apply and the Ward identity is preserved.  Without
vacuum polarization, the plus component of the dressed-electron
current is not renormalized~\protect\cite{BRS,ChiralLimit}.}

The analog of these difficulties with truncation can be induced
in Feynman perturbation theory by separating covariant
diagrams into time-ordered diagrams and discarding those time
orderings that include intermediate states with more particles
than some finite limit.  This destroys covariance, disrupts
regularization, and induces spectator dependence for
subdiagrams.  In the nonperturbative case, this happens
not just to some finite order in the coupling but to all
orders.

\section{Light-front coupled-cluster method}
\label{sec:Method}

To avoid truncation, we introduce the exponential-operator
technique of the many-body coupled-cluster (CC)
method~\cite{CCorigin,CCreviews}.
An eigenstate $|\psi(\ub{P})\rangle$ is written as 
$\sqrt{Z}e^T|\phi(\ub{P}\rangle$, where $|\phi(\ub{P})\rangle$
is limited to one or a few Fock sectors with the lowest
number(s) of constituents, the valence sector(s).  The operator $T$
is a sum of operators that increase particle number
but conserve momentum $\ub{P}$, the angular momentum
component\footnote{The other two components of angular
momentum are not kinematic~\protect\cite{DLCQreview}.
The eigenstates of the Hamiltonian
will in general be linear combinations of eigenstates of $J^2$.
Determination of the $J^2$ eigenstates is a separate dynamical
problem.} $J_z$, and all relevant 
quantum numbers, such as charge and baryon number.
The factor $\sqrt{Z}$ is a normalization factor, such that
$\langle\phi(\ub{P}')|\phi(\ub{P})\rangle
=\langle\psi(\ub{P}')|\psi(\ub{P})\rangle=\delta(\ub{P}'-\ub{P})$.
We then construct an effective eigenvalue problem in the valence
sector,
$P_v\ob{\Pminus}|\phi(\ub{P})\rangle
=\frac{M^2+P_\perp^2}{P^+}|\phi(\ub{P})\rangle$,
where $\ob{\Pminus}\equiv e^{-T}\Pminus e^T$ and $P_v$
is a projection onto the valence sector.
Equations for the functions that determine $T$ are found by
the orthogonal projection $(1-P_v)\ob{\Pminus}|\phi(\ub{P})\rangle=0$.
Up to this point, no approximation has been made, and the
problem remains infinite, because there are infinitely
many contributions to $T$.  

To have a finite set of equations, we
truncate $T$ to a few operators and truncate the projection
$1-P_v$ in a consistent way, such that just enough equations
are produced to be able to solve for the functions in the
truncated $T$ operator.  For example, if $T$ can create one
additional particle above the valence state, $1-P_v$ projects
onto only this additional Fock sector.  
After truncation, we have a finite
set of nonlinear equations for the functions in $T$, coupled
to the valence-sector wave functions, and a valence-sector
eigenvalue problem where the effective Hamiltonian depends
on the functions in $T$.  The former are essentially 
auxiliary equations that help define the latter.

What is not truncated is the
exponentiation of $T$, and thus the full Fock space can
be retained, though the wave functions for the higher
Fock sectors are clearly only approximate.
The effective Hamiltonian is computed from its 
Baker--Hausdorff expansion
$\ob{\Pminus}=\Pminus+[\Pminus,T]+\frac12[[\Pminus,T],T]+\cdots$.
Only a finite number of terms contributes, because each
factor of $T$ increases the number of particles created,
eventually exceeding the truncation of $(1-P_v)$.

Although this light-front coupled-cluster (LFCC) method
uses the mathematics of the traditional CC
method~\cite{CCorigin,CCreviews,CC-QFT}, it is quite different
conceptually.  In fact, the name coupled cluster does not
really apply, but we use it to acknowledge the origin
of the LFCC method.  The
CC method is applied to a single Fock sector, with a 
large number of constituents.  The $T$ operator builds
correlated excitations onto a Hartree--Fock-type
ground state.  Within products of $T$ there are
no contractions, because every term in $T$ annihilates
one or more of the single-particle states in the ground
state and creates one or more excited states.  In the
LFCC method, the valence sector has a small number of
constituents, and the method of solution of the 
eigenvalue problem here is left unspecified.  The
terms of the $T$ operator do include annihilation,
because the positive light-front momentum $P^+$
cannot be conserved unless one or more particles
are annihilated to provide momentum for those that
are created.  As a consequence, powers of $T$
include contractions, but these are needed in
order that $T$ to some power not annihilate
the entire valence state, which would effectively
truncate the exponentiation of $T$.

In addition to the fundamental mass eigenvalue problem,
the LFCC method must also contend with the evaluation
of matrix elements of operators, in order to be able
to extract physical quantities from the LFCC eigenstates.
This is nontrivial, because a direct calculation
of the normalizing factor $\sqrt{Z}$ is impractical,
due to the infinite set of terms in the sum over
Fock states within 
$\langle\phi(\ub{P})|e^{T^\dagger}e^T|\phi(\ub{P})\rangle$.
This same issue arises in the traditional CC method~\cite{CCreviews},
and there a technique exists for expectation values which
can be adapted for the LFCC method and extended to
include off-diagonal matrix elements.  Some care must be
taken, however, in that the LFCC method uses momentum
eigenstates with Dirac-delta normalization, unlike the
unit normalization of the standard CC states.  The 
normalization factor $\sqrt{Z}$ is introduced to avoid
division by 
$\langle\psi(\ub{P}')|\psi(\ub{P})\rangle=\delta(\ub{P}'-\ub{P})$
in the computation of expectation values.

For an operator $\hat{O}$ we write the expectation
value $\langle\hat{O}\rangle$ in the state
$\sqrt{Z}e^T|\phi(\ub{P})\rangle$ as
$\langle\hat{O}\rangle
=Z\langle\phi(\ub{P})|e^{T^\dagger}\hat{O}e^T|\phi(\ub{P})\rangle$
and define $\ob{O}=e^{-T}\hat{O}e^T$ and
\be 
\langle\widetilde\psi(\ub{P})|=Z\langle\phi(\ub{P})|e^{T^\dagger}e^T
=\sqrt{Z}\langle\psi(\ub{P})|e^T,
\ee
so that $\langle\hat{O}\rangle=\langle\widetilde\psi(\ub{P})|\ob{O}|\phi(\ub{P})\rangle$.
By construction, we have 
\be
\langle\widetilde\psi(\ub{P}')|\phi(\ub{P})\rangle
=\langle\psi(\ub{P}')|\psi(\ub{P})\rangle=\delta(\ub{P}'-\ub{P})
\ee
and
\be
\langle\widetilde\psi(\ub{P})|\ob{\Pminus}
=\sqrt{Z}\langle\psi(\ub{P})|e^T e^{-T}\Pminus e^T
=\frac{M^2+P_\perp^2}{P^+}\langle\widetilde\psi(\ub{P})|.
\ee
Thus, $\langle\widetilde\psi(\ub{P})|$ is a left eigenvector of the
(necessarily) non-Hermitian $\ob{\Pminus}$, with the same mass
eigenvalue, normalized such that the projection onto the 
valence state is a simple momentum-conserving delta function.
Therefore, an expectation value is computed by constructing
the effective operator $\ob{O}$ from a Baker--Hausdorff expansion,
solving the left-hand eigenvalue problem, and evaluating
the inner product $\langle\widetilde\psi(\ub{P})|\ob{O}|\phi(\ub{P})\rangle$.
As for $\ob{\Pminus}$, only a finite number of terms in the
Baker--Hausdorff expansion of $\ob{O}$ will contribute.
The extension to off-diagonal matrix elements is straightforward.

The left-hand eigenvalue problem must be truncated to an extent
consistent with the truncation of $T$, such that $\langle\widetilde\psi(\ub{P})|$ 
is limited to the Fock sectors of the valence state plus those 
created by application of $T$.  To understand the truncation,
consider the following.  Define an operator $L=(1-P_v)Ze^{T^\dagger}e^T P_\phi$,
with $P_\phi$ the projection onto the valence eigenstate $|\phi(\ub{P})\rangle$.
Because of the projection operators, $e^L$ is simply $1+L$.  The
left-hand eigenstate can then be written as
$\langle\widetilde\psi(\ub{P})|
=\langle\phi(\ub{P})|e^{L^\dagger}
 +Z\langle\phi(\ub{P})|e^{T^\dagger}e^T(P_v-P_\phi)^\dagger$.
We see, then, that $L$ plays the role of $T$, and therefore should 
be truncated in the same way.  The truncated left-hand eigenvalue problem 
creates a finite set of linear equations for the wave functions of
$\langle\widetilde\psi(\ub{P})|$.

\section{Model application}
\label{sec:model}

\begin{figure}[ht]
\vspace{0.2in}
\centerline{\includegraphics[width=12cm]{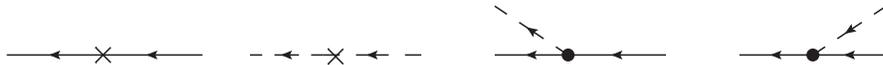}}
\caption{\label{fig:Fig2} Graphical representation
of the model Hamiltonian operator $\Pminus$ defined
in Eq.~(\ref{eq:Pminus}) of the text.
Each graph represents an operator that annihilates
one or more particles on the right and creates
one or more to take their place.  The crosses refer
to light-front kinetic-energy terms.}
\end{figure}

To illustrate the method, we apply it to a simple model where
an analytic solution is known~\cite{PV1}.  The model is a
light-front analog of the Greenberg--Schweber model~\cite{Greenberg}
for a static fermionic source that emits and absorbs bosons
without changing its spin.  In updated notation, the 
Hamiltonian given in \cite{PV1} can be written as
\bea \label{eq:Pminus}
\lefteqn{\Pminus = \int d\ub{p} \frac{M^2+M'_0 p^+}{P^+}\sum_s b_s^\dagger(\ub{p})b_s(\ub{p})
  +\int d\ub{q}\sum_l(-1)^l \frac{\mu_l^2+q_\perp^2}{q^+} a_l^\dagger(\ub{q})a_l(\ub{q})}&& \\
 &+\frac{g}{P^+} \int \frac{d\ub{p}d\ub{q}}{\sqrt{16\pi^3 q^+}}
  \sum_{ls}\left(\frac{p^+}{p^++q^+}\right)^\gamma 
  \left[a_l^\dagger(\ub{q})b_s^\dagger(\ub{q})b_s(\ub{p}+\ub{q})
    +b_s^\dagger(\ub{p}+\ub{q})b_s(\ub{p})a_l(\ub{q})\right], \nonumber
\eea
where $a_0^\dagger$ creates a ``physical'' boson of mass $\mu_0$,
$a_1^\dagger$ creates a Pauli--Villars (PV) boson of mass $\mu_1$,
and $b_s^\dagger$ creates the fermion with mass $M$ and spin $s$.
The parameter $\gamma$ can take any positive value; as shown in \cite{PV1},
it controls the longitudinal endpoint behavior of the wave functions.
The PV boson provides the necessary ultraviolet regularization, to
define the self-energy $M'_0$.  To accomplish the regularization,
the PV boson is assigned a negative norm.\footnote{In \protect\cite{PV1},
the PV cancellations were arranged by use of an imaginary coupling
rather than a negative norm.}  The (anti)commutation relations are
\be
\{b_x(\ub{p}),b_{s'}^\dagger(\ub{p}')\}=\delta_{ss'}\delta(\ub{p}-\ub{p}'), \;\;
[a_l(\ub{q}),a_{l'}^\dagger(\ub{q}')]=(-1)^l \delta_{ll'} \delta(\ub{q}-\ub{q}').
\ee
A graphical representation of the Hamiltonian
is given in Fig.~\ref{fig:Fig2}.  
The model is not fully covariant, which hides some of the power of
the LFCC method, but is sufficient to show how the method can be
applied.

\begin{figure}[ht]
\vspace{0.2in}
\centerline{\includegraphics[width=2.5cm]{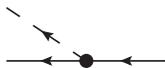}}
\caption{\label{fig:Fig3} Graphical representation
of the truncated $T$ operator.}
\end{figure}

\begin{figure}[ht]
\vspace{0.2in}
\begin{tabular}{c}
\centerline{\includegraphics[width=10cm]{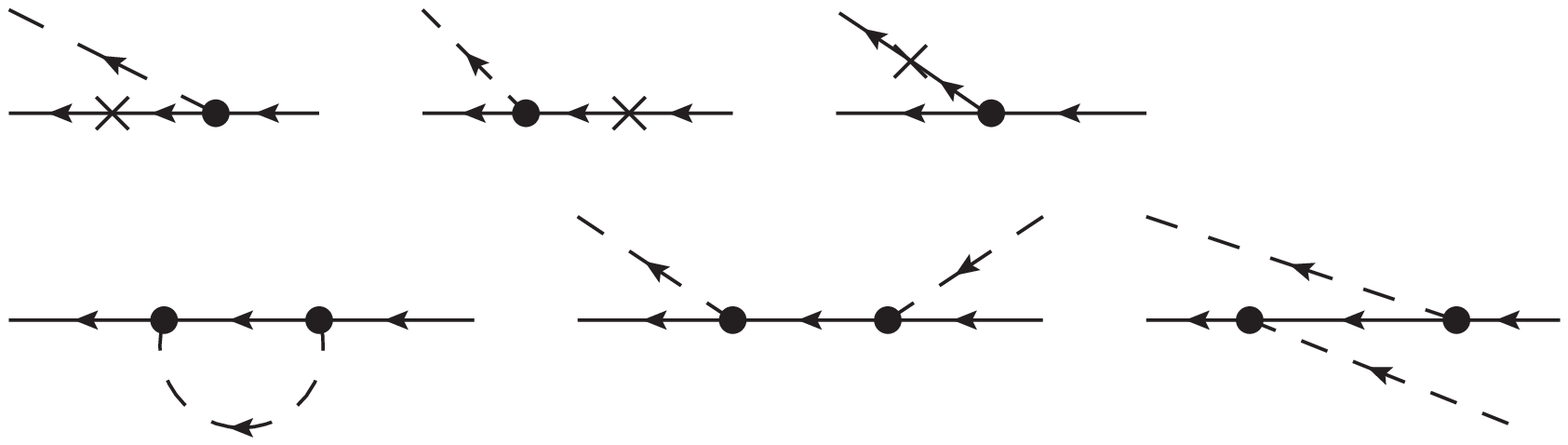}} \\
(a) \\
\mbox{ } \\
\centerline{\includegraphics[width=10cm]{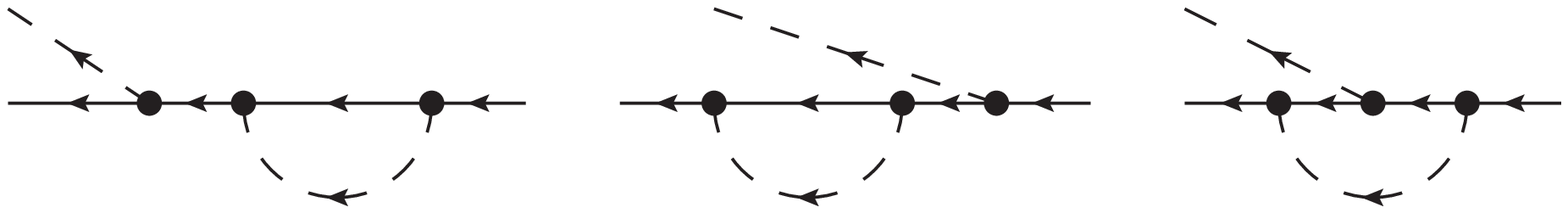}} \\
(b)
\end{tabular}
\caption{\label{fig:Fig4} Graphical representation
of the operators (a) $[\Pminus,T]$ and (b) $[[\Pminus,T],T]$.
The crosses indicate light-front kinetic-energy contributions.
The self-energy loops in the fourth diagram of (a) and the first
and second diagrams of (b) make identical contributions,
with no Fock-sector or spectator dependence.}
\end{figure}

We truncate the $T$ operator to include only boson emission
from the fermion, as represented in Fig.~\ref{fig:Fig3},
\be
T=\sum_{ls}\int d\ub{q} d\ub{p} \, t_{ls}(\ub{q},\ub{p})
   a_l^\dagger(\ub{q})b_s^\dagger(\ub{p})b_s(\ub{p}+\ub{q}),
\ee
with $t_{ls}$ the operator functions
to be determined.  
The effective Hamiltonian $\ob{\Pminus}$ is constructed from its
Baker--Hausdorff expansion.  A graphical representation of the
first two commutators is given in Fig.~\ref{fig:Fig4}.  
The expression for $\ob{\Pminus}$ is
\bea
\lefteqn{\ob{\Pminus}=
  \int d\ub{q}\sum_l (-1)^l \frac{\mu_l^2+q_\perp^2}{q^+}
                a_l^\dagger(\ub{q})a_l(\ub{q})} &&  \\
&&+\int d\ub{p} \sum_s b_s^\dagger(\ub{p})b_s(\ub{p}) 
              \left[ \frac{M^2+M'_0 p^+}{P^+} \right. \nonumber \\
&&  \rule{0.4in}{0mm} \left.
         +\frac{g}{P^+}\sum_l (-1)^l\int \frac{d\ub{q}}{\sqrt{16\pi^3 q^+}}
      \left(\frac{p^+-q^+}{p^+}\right)^\gamma 
          \theta(p^+-q^+)t_{ls}(\ub{q},\ub{p}-\ub{q})\right]  \nonumber \\
&& +\frac{g}{P^+}\int\frac{d\ub{p}d\ub{q}}{\sqrt{16\pi^3 q^+}}
           \sum_{ls}\left(\frac{p^+}{p^++q^+}\right)^\gamma
              b_s^\dagger(\ub{p}+\ub{q})b_s(\ub{p})a_l(\ub{q}) \nonumber \\
&& +\int d\ub{p} d\ub{q} \sum_{ls} a_l^\dagger(\ub{q})b_s^\dagger(\ub{p})
                                              b_s(\ub{p}+\ub{q}) \nonumber \\
&& \times\left\{\frac{g}{P^+}\frac{1}{\sqrt{16\pi^3 q^+}}
                        \left(\frac{p^+}{p^++q^+}\right)^\gamma
   +\left(\frac{\mu_l^2+q_\perp^2}{q^+}-\frac{M'_0q+}{P^+}\right)
                              t_{ls}(\ub{q},\ub{p})\right. \nonumber \\
&&  +\frac{g}{2P^+}\int\frac{d\ub{q}'}{\sqrt{16\pi^3 q^{\prime +}}}
\sum_{l'} (-1)^{l'}\left[\theta(p^+-q^{\prime +})
   \left(\frac{p^+-q^{\prime +}}{p^+}\right)^\gamma \right. \nonumber \\
&& \times\left\{t_{ls}(\ub{q},\ub{p})t_{l's}(\ub{q}',\ub{p}-\ub{q}')
+\theta(p^++q^+-q^{\prime +})
    t_{ls}(\ub{q},\ub{p}-\ub{q}')t_{l's}(\ub{q}',\ub{p}+\ub{q}-\ub{q}')\right\} \nonumber \\
&& \left. \left.
-2\theta(p^++q^+-q^{\prime +})\left(\frac{p^++q^+-q^{\prime +}}{p^++q^+}\right)^\gamma
         t_{ls}(\ub{q},\ub{p}) t_{l's}(\ub{q}',\ub{p}+\ub{q}-\ub{q}')\right]\right\}
         \nonumber \\
&& +\frac{g}{P^+}\int \frac{d\ub{p} d\ub{q} d\ub{q}'}{\sqrt{16\pi^3 q^+}}
\theta(p^++q^+-q^{\prime +})
\sum_{ll's} a_{l'}^\dagger(\ub{q}')b_s^\dagger(\ub{p}+\ub{q}-\ub{q}')b_s(\ub{p})a_l(\ub{q})
  \nonumber \\
&& \rule{0.5in}{0mm} \times\left[\theta(p^+-q^{\prime +})
   \left(\frac{p^+-q^{\prime +}}{p^++q^+-q^{\prime +}}\right)^\gamma
                 t_{l's}(\ub{q}',\ub{p}-\ub{q}')  \right. \nonumber \\
&& \rule{0.7in}{0mm} \left.
  -\left(\frac{p^+}{p^++q^+}\right)^\gamma t_{l's}(\ub{q}',\ub{p}+\ub{q}-\ub{q}')\right],
                   \nonumber
\eea
where we list only terms that connect the lowest Fock sectors. 
Notice that the self-energy contribution $M'_0$ is the same
in all Fock sectors and that Fig.~\ref{fig:Fig4}(b)
contains all three of the diagrams analogous to
those for the Ward identity in QED, as discussed in the
Introduction, with no truncation in particle number.

The valence state is the bare-fermion state
$|\phi^\sigma(\ub{P})\rangle=b_\sigma^\dagger(\ub{P})|0\rangle$.
The projection $1-P_v$ is
truncated to the one-fermion/one-boson sector.
The full eigenstate is
$|\psi^\sigma(\ub{P})\rangle=\sqrt{Z}e^T|\phi^\sigma(\ub{P})\rangle$,
where we have generalized the basic construction to have one $T$
operator for both spins $\sigma=\pm$ and
will solve for both states simultaneously.  This allows
contractions in powers of $T$ to include sums over both spins.
The truncated left-hand eigenvector is
\be
\langle\widetilde\psi^\sigma(\ub{P})|
       =\langle\phi^\sigma(\ub{P})|
+\sum_{ls}\int d\ub{q}\theta(P^+-q^+)
l_{ls}^{\sigma*}(\ub{q},\ub{P})\langle0|a_l(\ub{q}) 
b_s(\ub{P}-\ub{q}),
\ee
where $l_{ls}^\sigma$ is the 
left-hand one-fermion/one-boson wave function.
Due to the lack of covariance in the model, these states are all limited
to having a fixed total transverse momentum $\vec P_\perp$, which we
take to be zero.

The eigenvalue problem in the valence sector 
$P_v\ob{\Pminus}|\phi^\sigma(\ub{P})\rangle
=\frac{M^2}{P^+}|\phi^\sigma(\ub{P})\rangle$
becomes
\bea
\lefteqn{\frac{M^2+M'_0 P^+}{P^+}|\phi^\pm(\ub{P})\rangle}&& \\
&& +\frac{g}{P^+}\int\frac{d\ub{q}}{\sqrt{16\pi^3 q^+}}
\theta(P^+-q^+)\left(\frac{P^+-q^+}{P^+}\right)^\gamma 
     \sum_l (-1)^l t_{l\pm}(\ub{q},\ub{P}-\ub{q})|\phi^\pm(\ub{P})\rangle \nonumber \\
 &&    =\frac{M^2}{P^+}|\phi^\pm(\ub{P})\rangle, \nonumber
\eea
which reduces to a determination of the self-energy
\be \label{eq:selfenergy}
M'_0=-\frac{g}{P^+}\int\frac{d\ub{q}}{\sqrt{16\pi^3 q^+}}
\theta(P^+-q^+)\left(\frac{P^+-q^+}{P^+}\right)^\gamma 
     \sum_l (-1)^l t_{l\pm}(\ub{q},\ub{P}-\ub{q}).
\ee
In the one-fermion/one-boson sector, we have 
$(1-P_v)\ob{\Pminus}|\phi^\pm(\ub{P})\rangle=0$ or
\bea \label{eq:nonvalence}
\lefteqn{\int d\ub{q} \theta(P^+-q^+)
\sum_l a_l^\dagger(\ub{q})b_\pm^\dagger(\ub{P}-\ub{q})|0\rangle}&& \\
&& \times\left\{\frac{g}{P^+}\frac{1}{\sqrt{16\pi^3 q^+}}
                        \left(\frac{P^+-q^+}{P^+}\right)^\gamma
   +\left(\frac{\mu_l^2+q_\perp^2}{q^+}-\frac{M'_0q+}{P^+}\right)
                              t_{ls}(\ub{q},\ub{P}-\ub{q})\right. \nonumber \\
&&  +\frac{g}{2P^+}\int\frac{d\ub{q}'}{\sqrt{16\pi^3 q^{\prime +}}}
\sum_{l'} (-1)^{l'}\left[\theta(P^+-q^+-q^{\prime +})
   \left(\frac{P^+-q^+-q^{\prime +}}{P^+-q^+}\right)^\gamma \right. \nonumber \\
&& \times\left\{t_{l\pm}(\ub{q},\ub{P}-\ub{q})t_{l'\pm}(\ub{q}',\ub{P}-\ub{q}-\ub{q}')
    \right. \nonumber \\
&& \left.  \rule{0.5in}{0mm}
+\theta(P^+-q^{\prime +})
    t_{l\pm}(\ub{q},\ub{P}-\ub{q}-\ub{q}')t_{l'\pm}(\ub{q}',\ub{P}-\ub{q}')\right\} \nonumber \\
&& \left. \left.
-2\theta(P^+-q^{\prime +})\left(\frac{P^+-q^{\prime +}}{P^+}\right)^\gamma
         t_{l\pm}(\ub{q},\ub{P}-\ub{q}) t_{l'\pm}(\ub{q}',\ub{P}-\ub{q}')\right]\right\}=0.
         \nonumber 
\eea
Thus, the contents of the outer curly brackets must sum to zero.
This will occur if the function $t_{ls}$ is 
\be
t_{ls}(\ub{q},\ub{p})=\frac{-g}{\sqrt{16\pi^3 q^+}}\left(\frac{p^+}{p^++q^+}\right)^\gamma
   \frac{q^+/P^+}{\mu_l^2+q_\perp^2}.
\ee
The fact that the self-energy $M'_0$ is the same in the valence
sector and the one-fermion/one-boson sector plays a critical role;
the expression (\ref{eq:selfenergy}) obtained in the valence sector
is exactly what is needed to obtain the necessary cancellations
in (\ref{eq:nonvalence}). 
The self-energy can be computed from Eq.~(\ref{eq:selfenergy}) as
\be
M'_0=\frac{g^2}{16\pi^3P^+}\frac{\ln(\mu_1/\mu_0)}{\gamma +1/2},
\ee
which agrees with the result in \cite{PV1}.  In fact, with
$t_{ls}$ as given above, the exponential operator $e^T$
generates the exact solution given in \cite{PV1}.

The solution for $t_{ls}$
provides the input to the left-hand eigenvalue problem, 
$\langle\widetilde\psi^\pm(\ub{P})|\ob{\Pminus}
=\frac{M^2}{P^+}\langle\widetilde\psi^\pm(\ub{P})|$.
The effective Hamiltonian $\ob{\Pminus}$ simplifies considerably;
the square bracket in the $b^\dagger b$ term becomes just $M^2/P^+$
and the entire $a^\dagger b^\dagger b$ term, which corresponds
to the curly brackets in (\ref{eq:nonvalence}), is zero.  
The remaining terms in $\ob{\Pminus}$ yield the following
integral equation for the left-hand wave function:
\bea
\lefteqn{\frac{g}{P^+}\frac{1}{\sqrt{16\pi^3 q^+}}
      \left(\frac{P^+-q^+}{P^+}\right)^\gamma \delta_{s\pm}
      +\frac{\mu_l^2+q_\perp^2}{q^+} l_{ls}^\pm(\ub{q},\ub{P})}&& \\
&& -\left(\frac{g}{P^+}\right)^2 \int\frac{d\ub{q}'}{\sqrt{16\pi^3 q^+}}
      \theta(P^+-q^{\prime +})\sqrt{\frac{q^{\prime +}}{16 \pi^3}}
      \sum_{l'} (-1)^{l'} \frac{1}{\mu_{l'}^2+q_\perp^{\prime 2}}
          l_{l's}^\pm(\ub{q}',\ub{P})  \nonumber \\
&&   \rule{0.4in}{0mm} \times\left[\theta(P^+-q^+-q^{\prime +})
            \frac{(P^+-q^+-q^{\prime +})^{2\gamma}}
                   {(P^+-q^{\prime +})^\gamma(P^+-q^+)^\gamma} \right. \nonumber \\
&& \rule{0.6in}{0mm} \left.
        -\left(\frac{P^+-q^+}{P^+}\right)^\gamma
        \left(\frac{P^+-q^{\prime +}}{P^+}\right)^\gamma \right]=0. \nonumber
\eea
Following the pattern of the inhomogeneous term, we can seek a solution of the form
\be
l_{ls}^\sigma(\ub{q},\ub{P})=\delta_{\sigma s}\frac{-g}{\sqrt{16\pi^3 q^+}}
   \left(\frac{P^+-q^+}{P^+}\right)^\gamma \frac{q^+/P^+}{\mu_l^2+q_\perp^2}\tilde{l}(q^+/P^+).
\ee
Substitution yields a one-dimensional integral equation for $\tilde{l}(y)$
\be
\tilde{l}(y)=1+\frac{g^2}{16\pi^2}\frac{\mu_1^2-\mu_0^2}{\mu_0^2\mu_1^2}
\int_0^1 dy' (1-y')^{2\gamma} y'[(1-y)^2\tilde{l}(y'(1-y))-\tilde{l}(y')].
\ee
This equation can be solved iteratively, to generate an expansion in 
powers of $g^2$, or numerically.  A Gauss--Jacobi quadrature will
convert the integral equation into a linear system for the values
of $\tilde{l}$ at the chosen quadrature points.
The solution then provides the rest of the 
information needed for the computation of matrix elements.

To consider a particular matrix element as an example, we
compute the Dirac form factor for the dressed fermion from
a matrix element of the current 
$J^+=\ob{\psi}\gamma^+\psi$.  The current
couples to a photon of momentum $q$.  With our normalization,
the matrix element is generally~\cite{BrodskyDrell}
\be
\langle\psi^\sigma(\ub{P}+\ub{q})|16\pi^3J^+(0)|\psi^\pm(\ub{P})\rangle
=2\delta_{\sigma\pm}F_1(q^2)\pm\frac{q^1\pm iq^2}{M}\delta_{\sigma\mp}F_2(q^2),
\ee
with $F_1$ and $F_2$ the Dirac and Pauli form factors.
In the present model, the fermion cannot flip its spin; therefore,
$F_2$ is zero, and we investigate only $F_1$.

In the LFCC method, the form factor is given by the matrix element
\be
F_1(q^2)=8\pi^3\langle\widetilde\psi^\pm(\ub{P}+\ub{q})|\ob{J^+(0)}|\phi^\pm(\ub{P})\rangle,
\ee
with $\ob{J^+(0)}=J^+(0)+[J^+(0),T]+\cdots$.  For this model, there
are no contributions from fermion-antifermion pairs, so that
\be
J^+(0)=2\sum_s\int\frac{d\ub{p}'}{\sqrt{16\pi^3}}\int\frac{d\ub{p}}{\sqrt{16\pi^3}}
    b_s^\dagger(\ub{p}')b_s(\ub{p}),
\ee
and only the first two terms of the Baker--Hausdorff expansion contribute
to the matrix element.  The second term is
\be
[J^+(0),T]=2\sum_{ls}\int\frac{d\ub{p}'}{\sqrt{16\pi^3}}\int\frac{d\ub{p}}{\sqrt{16\pi^3}}
  \int d\ub{q}'   [t_{ls}(\ub{q}',\ub{p})-t_{ls}(\ub{q}',\ub{p}']
  a_l^\dagger(\ub{q}') b_s^\dagger(\ub{p}')b_s(\ub{p}).
\ee
The first term contributes $1/8\pi^3$ to the matrix element; the
second contributes
\bea
\lefteqn{\langle\widetilde\psi^\pm(\ub{P}+\ub{q})|[J^+(0),T]|\phi^\pm(\ub{P})\rangle
=\frac{1}{8\pi^3}\sum_{l}(-1)^l \int d\ub{q}'\theta(P^++q^+-q^{\prime +})} && \nonumber \\
&& \times l_{l\pm}^\pm(\ub{q}',\ub{P}+\ub{q})
  [\theta(P^+-q^{\prime +})t_{l\pm}(\ub{q}',\ub{P}-\ub{q}')
    -t_{l\pm}(\ub{q}',\ub{P}+\ub{q}-\ub{q}')].
\eea

Because the model limits calculations to a fixed total transverse momentum,
we calculate the matrix element in a frame where $\vec q_\perp$ is zero and
$q^+$ is not.\footnote{In \protect\cite{PV1} the matrix element was computed
in a frame where $q^+=0$.  This could be done because the wave functions
of the exact solution were taken to be boost invariant.  Here, although the
right-hand eigenvalue problem has accidentally provided the exact solution,
we continue with the LFCC approximation in the calculation of the matrix
element, to provide a more complete illustration of the method.}
With $\alpha\equiv q^+/P^+$ and $P'=P+q$, we have
\bea \label{eq:alpha}
q^2&=&(P'-P)^2=2M^2-P^{\prime +}P^--P^{\prime -}P^+ 
     \nonumber \\
   &=&2M^2-M^2(1+\alpha)-\frac{M^2}{1+\alpha}=-\frac{M^2\alpha^2}{1+\alpha}.
\eea
On substitution of the solutions
for the wave functions and evaluation of the transverse integral,
the form factor can be written as a function of $\alpha$
\bea \label{eq:formfactor}
F_1(q^2)&=&1+\frac{g^2}{16\pi^2}(1+\alpha)\frac{\mu_1^2-\mu_0^2}{\mu_0^2\mu_1^2}
\left[\int_0^{1/(1+\alpha)} dy\,\tilde{l}(y) y(1-y)^\gamma[1-(1+\alpha)y]^\gamma
                     \right. \nonumber \\
&& \rule{1.5in}{0mm} \left.
-\int_0^1 dy\,\tilde{l}(y) y(1-y)^{2\gamma}\right]. 
\eea
The PV dependence is easily removed in the limit of an infinite PV mass
($\mu_1\rightarrow\infty$).  If $\tilde{l}$ is computed in quadrature,
the integrals remaining in $F_1$ can be computed from the same quadrature
rule for any chosen value of $\alpha$.  If $\tilde{l}$ is instead
constructed as an expansion in $g^2$, $F_1$ can also be constructed
as an expansion.  In any case, in the limit of $q^2\rightarrow0$,
we have from (\ref{eq:alpha}) that $\alpha=0$ and, because the two
integrals in (\ref{eq:formfactor}) then cancel, $F_1(0)=1$,
consistent with the unit charge in the current $J^+=\bar\psi\gamma^+\psi$.

\section{Summary}
\label{sec:Summary}

We have proposed a new Hamiltonian method for the nonperturbative
solution of quantum field theories that avoids Fock-space truncations.
The full eigenstate is constructed from the action of an exponentiated
operator $T$ on a valence state $|\phi\rangle$.  This yields a
valence eigenvalue problem 
$\ob{\Pminus}|\phi\rangle=\frac{M^2+P_\perp^2}{P^+}|\phi\rangle$,
with $\ob{\Pminus}=e^{-T}\Pminus e^T$, and a set of auxiliary 
equations for the functions in $T$.  Expectation values are
computed as 
$\langle\hat{O}\rangle=\langle\widetilde\psi|e^{-T}\hat{O}e^T|\phi\rangle$,
with use of the left-hand eigenstate $\langle\widetilde\psi|$.  The
method then generates approximations by truncation of $T$ rather 
than of Fock space.

The application to the simple model in Sec.~\ref{sec:model}
shows that the construction of $\ob{\Pminus}$ generates self-energy
contributions that are Fock-sector and spectator independent.  Thus,
the uncanceled divergences that can arise from Fock-space truncations
do not occur; the self-energy is the same in every sector.  The 
application also shows that a simple approximation for the $T$ 
operator can provide a very good approximation to the eigenstate;
in this special case, the eigenstate is exact.  The calculation
of a matrix element is demonstrated in the calculation of a
Dirac form factor.

The LFCC method is not limited to any particular theory or
model, nor to Pauli--Villars regularization.  It should be
applicable to any regularized field theory.  
Work on an application to QED is in progress, with some
preliminary discussion given in~\cite{LC11}.
For theories with symmetry breaking and
vacuum structure, modes of zero longitudinal momentum~\cite{ZeroModes}
play some role and would require extension of the method to
include them; in particular, the contributions to the $T$
operator would not require annihilation, and the exponentiation
would produce generalized coherent states.  For discrete 
light-cone quantization (DLCQ)~\cite{PauliBrodsky,DLCQreview},
where longitudinal momentum fractions are restricted to integer
multiples of a fundamental amount $1/K$, truncation to $K$
particles is automatic; however, the method could still be
applied as a way of reducing the effective dimension of the
underlying matrix eigenvalue problem and allowing
higher resolution.  Even the supersymmetric form
of DLCQ (SDLCQ)~\cite{SDLCQ} should be amenable; instead
of constructing $\Pminus$ from a discretized supercharge
$Q^-$ via $\Pminus=\{Q^-,Q^-\}/2\sqrt{2}$, to retain the
supersymmetric spectrum, the effective Hamiltonian
$\ob{\Pminus}$ would be constructed from effective 
supercharges $\ob{Q^-}\equiv e^{-T}Q^-e^T$.  Thus, there
is a wide range of applications to consider.

\acknowledgments
This work was supported in part by the U.S. Department of Energy
through Contract No.\ DE-FG02-98ER41087.


\end{document}